\documentclass[preprint]{acmart}

\usepackage{multirow}
\usepackage{tabularx}
\usepackage{mathtools}
\usepackage{adjustbox}
\usepackage{dblfloatfix}
\usepackage{algorithm}
\usepackage{algpseudocode}
\AtBeginDocument{%
  }

\setcopyright{acmlicensed}
\copyrightyear{2025}
\acmYear{2025}
\acmDOI{XXXXXXX.XXXXXXX}

\begin{document}

\title{FLAME: Flexible LLM-Assisted Moderation Engine}


\author{Ivan Bakulin}
\authornote{Co-first authors.}
\orcid{0000-0003-4733-2275}
\author{Ilia Kopanichuk}
\authornotemark[1]
\email{kopanichuk@airi.net}
\affiliation{%
  \institution{AIRI, Moscow Institute of Physics and Technology}
  \city{Moscow}
  \country{Russia}
}

\author{Iaroslav Bespalov}
\affiliation{%
    \institution{AIRI}
    \city{Moscow}
    \country{Russia}
}

\author{Nikita Radchenko}
\affiliation{%
  \institution{SberHealth}
  \city{Moscow}
  \country{Russia}
}

\author{Vladimir Shaposhnikov}
\author{Dmitry V. Dylov}
\authornote{Co-senior authors.}
\author{Ivan Oseledets}
\authornotemark[2]
\affiliation{%
    \institution{AIRI, Skolkovo Institute of Science and Technology}
    \city{Moscow}
    \country{Russia}
}


\begin{abstract}
The rapid advancement of Large Language Models (LLMs) has introduced significant challenges in moderating user-model interactions. While LLMs demonstrate remarkable capabilities, they remain vulnerable to adversarial attacks, particularly ``jailbreaking'' techniques that bypass content safety measures. Current content moderation systems, which primarily rely on input prompt filtering, have proven insufficient, with techniques like Best-of-N (BoN) jailbreaking achieving success rates of 80\% or more against popular LLMs.

In this paper, we introduce Flexible LLM-Assisted Moderation Engine (FLAME): a new approach that shifts the focus from input filtering to output moderation. Unlike traditional circuit-breaking methods that analyze user queries, FLAME evaluates model responses, offering several key advantages: (1) computational efficiency in both training and inference, (2) enhanced resistance to BoN jailbreaking attacks, and (3) flexibility in defining and updating safety criteria through customizable topic filtering.
Our experiments demonstrate that FLAME significantly outperforms current moderation systems. For example, FLAME reduces attack success rate in GPT-4o-mini and DeepSeek-v3 by a factor of $\sim$9, while maintaining low computational overhead. We provide comprehensive evaluation on various LLMs and analyze the engine's efficiency against the state-of-the-art jailbreaking. This work contributes to the development of more robust and adaptable content moderation systems for LLMs.
\end{abstract}





\maketitle
\section{Introduction} 



Content moderation for Large Language Models (LLMs) represents a critical challenge in ensuring safe and appropriate human-AI interactions. While traditional content moderation approaches have focused primarily on input filtering \cite{huang2024contentmoderationllmaccuracy}, the increasing sophistication of adversarial techniques necessitates a fundamental shift in how we approach this problem.
The main purpose of content moderation is to limit the processing of requests that do not align with the system's intended use. This serves several crucial functions: prevents errors in specialized applications like medical consultation, eliminates irrelevant or potentially harmful data that may interfere with model performance, and reduces the risks associated with malicious information or legal violations \cite{chua2024aisafetygenerativeai}. Current moderation systems typically rely on input filtering to identify and block potentially harmful queries before they reach the model.

However, recent developments in jailbreaking techniques, particularly the Best-of-N (BoN) approach, have exposed significant vulnerabilities in existing moderation systems. These techniques exploit the probabilistic nature of LLM outputs through multiple sampling attempts, achieving concerning success rates of 80\% or higher against popular models \cite{hughes2024bestofnjailbreaking}. The effectiveness of these attacks highlights a critical gap in current defensive approaches, which predominantly focus on analyzing user inputs rather than model outputs. 

Our work introduces FLAME (Flexible LLM-Assisted Moderation Engine), \emph{shifting the focus from input filtering to output moderation} through an efficient regulatory policy. Unlike existing solutions that require extensive computational resources or complex neural architectures, FLAME employs a lightweight approach that can be deployed with minimal requirements. This design choice not only makes the system more accessible but also enables rapid adaptation to emerging threats through customizable topic filtering.
The flexibility of FLAME's architecture addresses one key limitation in current moderation systems: the ability to quickly adapt to new types of harmful content while maintaining efficient operation. Our approach allows organizations to define and update their moderation criteria based on specific needs and emerging challenges, without requiring significant retraining or computational resources.

Recent work on constitutional AI and classifier-based approaches \cite{sharma2025constitutionalclassifiers} has demonstrated the potential of sophisticated moderation systems. However, these solutions often demand substantial computational resources for both training and inference. In contrast, FLAME demonstrates that effective moderation can be achieved through carefully designed rule-based systems enhanced by LLM-generated training data. Through extensive experimentation and real-world deployment, we have validated this approach across multiple leading LLM platforms, consistently achieving a 2-9$\times$ improvement in resistance to BoN attacks. \\

\textbf{Contributions} This work advances the field of LLM content moderation in several ways. 
\begin{itemize}
    \item We introduce an \textit{output-centered }moderation approach that provides superior protection against state-of-the-art jailbreaking techniques while maintaining minimal computational requirements.
    \item We jailbreak 6 popular LLMs with and without our moderation engine, demonstrating an \textit{up to 9-fold improvement in their resistance to adversarial attacks}.
    \item Our engine challenges the prevailing trend towards resource-intensive censorship, demonstrating that effective moderation can be achieved without extensive model fine-tuning or complex neural architectures. 
    \item We report practical insights from deployment of the moderation engine into a dialogue system product, addressing critical considerations from the standpoint of user experience and finding the delicate balance between moderation strictness and system accessibility.
\end{itemize}

\section{Related work} 

\textbf{Classifier guards.} Markov et al.~\cite{markov2023holisticapproachundesiredcontent} proposed an active learning strategy that identifies relevant samples for labeling, balances between uncertainty and diversity, and leverages redundancy to capture rare events. Rebedea et al.~\cite{rebedea2023nemoguardrailstoolkitcontrollable} developed guardrails control LLM output, preventing harmful topics, following dialogue paths, and maintaining language styles. Chi et al.~\cite{chi2024llamaguard3vision} introduced multimodal LLM-based safeguard that classifies content as safe or unsafe based on user-provided guidelines, conversation context, and output formats. Kim et al.~\cite{kim2024testinglimitsjailbreakingdefenses} highlight the necessity of developing better definitions for unsafe outputs. Wang et al.~\cite{wang2024jailbreakdefensenarrowdomain} point to challenges in jailbreak defense even in narrow contexts, suggests future directions involving classifier calibration and human feedback integration, suggesting future directions involving classifier calibration and human feedback integration. Sharma et al.~\cite{sharma2025constitutionalclassifiersdefendinguniversal} proposed "Constitutional Classifiers", which use natural-language rules to train classifier safeguards defining what constitutes permitted and restricted content.

\textbf{Jailbreaking} Lapid et al.~\cite{lapid2024opensesameuniversalblack} developed a black-box Genetic Algorithm (GA) to manipulate LLMs. Huang et al.~\cite{huang2023catastrophicjailbreakopensourcellms} show that manipulation of generation strategies, such as removing system prompts and altering decoding parameters, can easily disrupt model alignment. Samvelyan et al.~\cite{samvelyan2024rainbowteamingopenendedgeneration} introduced "Rainbow Teaming" methodology that sefines features like risk category and attack style to diversify prompts and rank them based on their effectiveness. Doumbouya et al.~\cite{doumbouya2024h4rm3ldynamicbenchmarkcomposable} developed an open-source automated red-teaming platform for generating and analyzing jailbreak attacks. Andriushchenko et al.~\cite{andriushchenko2024jailbreakingleadingsafetyalignedllms} applied prompt templates, random suffix search, self transfer from easier tasks, transfer and prefilling attacks and showed that adaptive attacks are necessary to accurately assess LLM robustness. Hughes et al.~\cite{hughes2024bestofnjailbreaking} proposed "Best-of-N (BoN) Jailbreaking," a black-box algorithm and demonstrated its effectiveness across text, vision, and audio language models, achieving high Attack Success Rates (ASR).

\section{Method} 

\begin{algorithm}
\caption{\label{alg:flame}FLAME Training Pipeline}
\begin{algorithmic}[1]
    \Require
    \State topics - list of sensitive topics to filter
    \State training\_dialogs - collection of safe dialogues
    \Ensure
    \State blacklist - set of filtered phrases
    
    \Function{GenerateMessages}{topics}
        \State messages $\gets$ []
        \ForAll{topic in topics}
            \State // Generate variations using LLM
            \State variations $\gets$ \Call{GenerateTopicVariations}{topic, n=30}
            \State // Create semantic neighbors for robustness
            \State messages $\gets$ messages $\cup$ \Call{CreateSemanticMessages}{variations, k=20}
        \EndFor
        \State \Return messages
    \EndFunction
    
    \Function{AssembleTrainingCollection}{dialogs}
        \State training\_set $\gets$ []
        \ForAll{dialog in dialogs}
            \State training\_set $\gets$ training\_set $\cup$ \Call{DialogToMessages}{dialog}
        \EndFor
        \State \Return training\_set
    \EndFunction
    
    \Function{CreateBlacklist}{messages, training\_set}
        \State // Initial blacklist generation
        \State raw\_blacklist $\gets$ \Call{GenerateInitialBlacklist}{messages}
        
        \State // Clean and optimize blacklist
        \State refined\_blacklist $\gets \{\}$
        \ForAll{phrase in raw\_blacklist}
            \If{\Call{ValidateOnTrainingSet}{phrase, training\_set}}
                \State refined\_blacklist.add(phrase)
            \EndIf
        \EndFor
        
        \State \Return refined\_blacklist
    \EndFunction
    
    \State // Main execution pipeline
    \State messages $\gets$ \Call{GenerateMessages}{topics}
    \State training\_data $\gets$ \Call{AssembleTrainingCollection}{training\_dialogs}
    \State blacklist $\gets$ \Call{CreateBlacklist}{messages, training\_data}

\end{algorithmic}
\end{algorithm}

The moderation Algorithm \ref{alg:flame} checks whether a user request or a model response contain a banned topic. It is a binary classification problem:  0 --- no banned topics, 1 --- contains banned topic. \\
The inference algorithm itself is rule-based and relatively simple. Message (text string) $r$ is split into several multisets of $n$-grams of normal word forms, by the split function $S(r, n)$. Then, the classifying function $\sigma$ is computed as follows:
\begin{equation}
\sigma(r) = 
\begin{dcases}
        1, & \texttt{if} \sum_{n=1}^{k} \chi(S(r, n), T) > 0 \\
        0, & \texttt{if} \sum_{n=1}^{k} \chi(S(r, n), T) = 0
    \end{dcases}
\end{equation}
where $\chi$ is the characteristic function computed by the exact match of items,  $T$ is the set of all banned $n$-grams (\textit{i.e.}, \texttt{blacklist} in Algorithm \ref{alg:flame}), $k = 3$ is the maximum $n$-gram size. We used \textit{pymorphy3} \cite{pymorphy3} to normalize word forms and \textit{nltk} \cite{nltk} for the $n$-gram partitioning.  An example of a message split from a medical dialogue is shown below:

\begin{equation}
S(\texttt{My stomach hurts}, 2) = [\![\texttt{i stomach, stomach hurt} ]\!]
\end{equation}

Before discussing the details, we want to note that the engine itself is flexible. Topic selection in the example above allows for skipping domain-irrelevant topics while filtering out dangerous or forbidden subjects. One can choose any sets of such topics when adapting the engine to their needs.

\subsection{Assembly of set of banned $n$-grams}

The proposed method for creating a dataset for text classification is similar to the constitutional classifier created in \cite{sharma2025constitutionalclassifiers}. Unlike the constitutional classifier, based on LLM fine-tuning, FLAME works on classical methods for matching normalized forms of $n$-grams, so it is much more efficient in both training and inference. \textbf{Our method requires no GPUs for training.} One merely needs an LLM API (without moderation) and 1 CPU to get through a full training cycle (Algorithm \ref{alg:flame}) in a few hours.

The assembly of a set $T$ of banned $n$-grams consists of several sequential steps:

\begin{enumerate}
    \item topics selection,
    \item messages generation,
    \item messages preprocessing,
    \item $n$-grams filtration.
\end{enumerate}

\begin{figure}[htbp]
    \centering
    \includegraphics[width=0.5\textwidth]{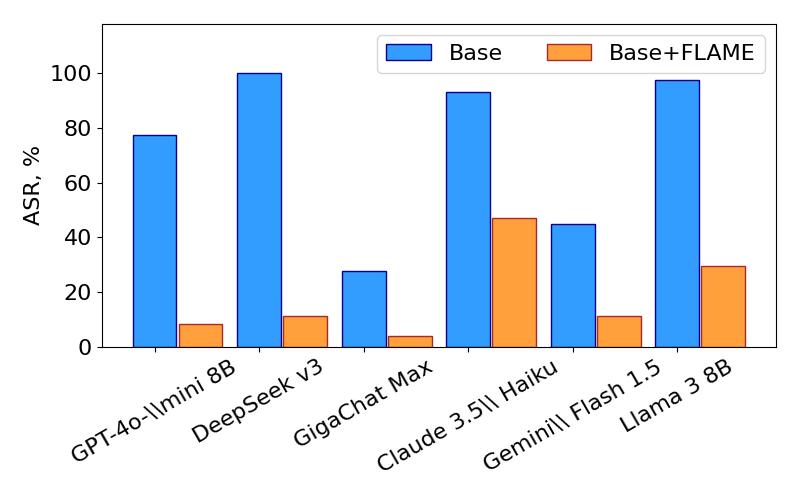}
    \caption{Attack success rate of BoN jailbreaks \cite{hughes2024bestofnjailbreaking} on moderation systems of different LLMs, showcasing the resilience that FLAME adds to popular LLMs.}
    \label{fig:asr_final}
\end{figure}

The first step was to select the topics we want to avoid in the model response. Once the set of topics has been established, each topic is matched with a set of user queries. Each query contains jailbreak attempt and used as an example for generation in the next step. We use the following criteria to select the topics and queries for them:

\begin{itemize}
    \item Topics related to activities that violate the law or international rights: terrorism, extremism, violence, {\it{etc.}}
    \item The system is focused on helping with medical advice, so topics outside of domain (e.g., esotericism, cooking, programming) have been added to the list. 
    \item Topics that may cause social stigma or inconsistency with the goals of the system are included. For example, bans on political discussions are associated with increased social sensitivity and legal restrictions
\end{itemize}

The main challenge was to strike a balance between filtering and usefulness. For example, words such as “alcohol” or “drug” have both medical relevance and irrelevant context.

\begin{figure*}[hbtp]
    \centering
    \includegraphics[width=\textwidth]{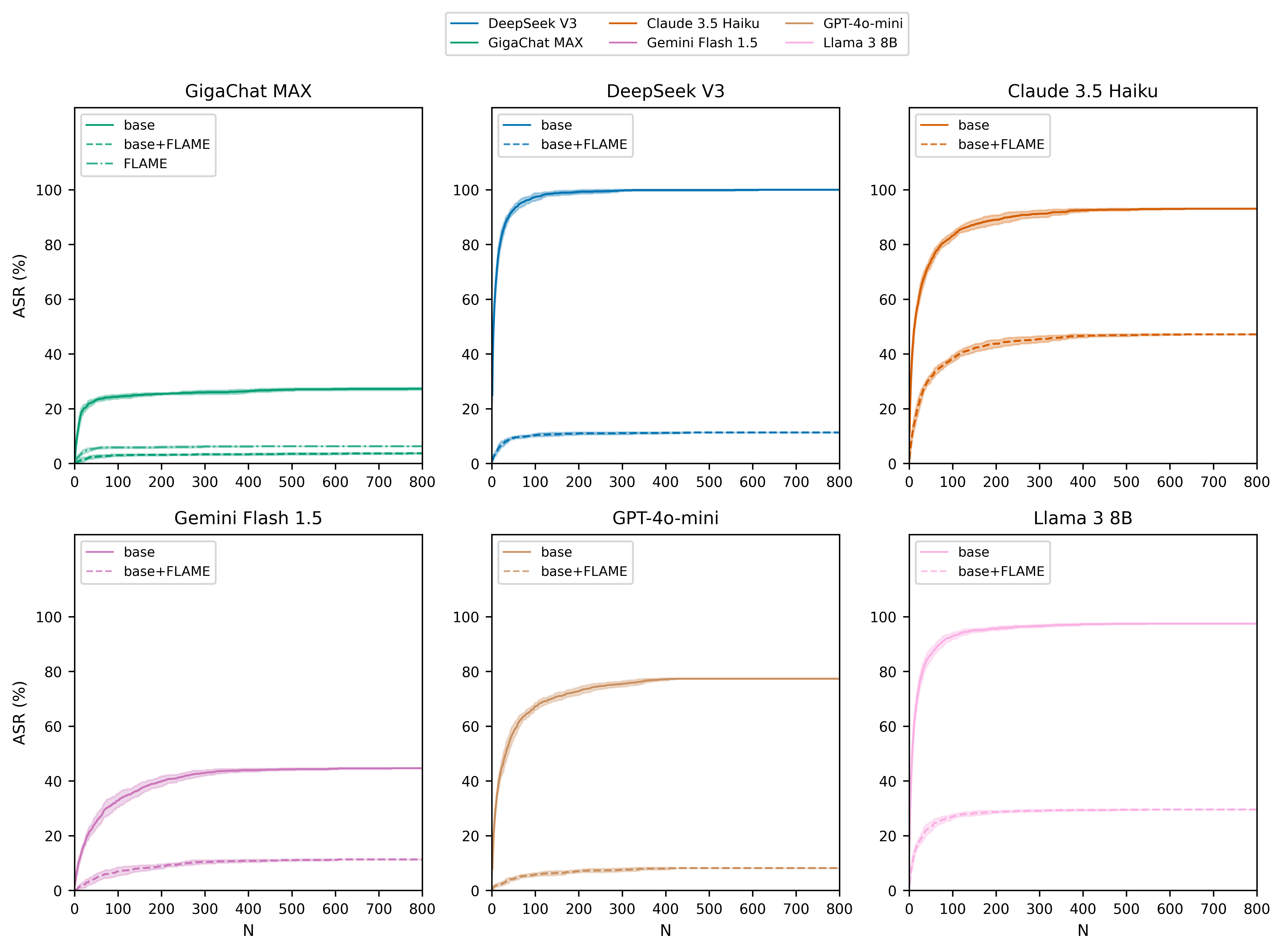}
    \caption{The dependence of attack success rate on the number of attempts by BoN for different LLMs compared to FLAME}
    \label{fig:fig1}
\vspace{-0.5cm}
\end{figure*}

In the second step, we generated a set of messages $\mathfrak{C}$ for each topic using the unmoderated LLM API by GigaChat Max. We used pre-made examples and prompts with jailbreak attempts from the previous step.

In the third step, we split each message $C$ into multiset of $n$-grams of normal word forms as $W(C, n), n \in \{1,...,k\}$, $k=3$ is the maximum $n$-gram size. We formed the union multiset of all generated $n$-grams: 
\begin{equation}
\mathfrak{T} = \bigcup_{j=1}^{|\mathfrak{C}|} \bigcup_{n=1}^{k}  W(C_j, n)    
\end{equation}

The most important part of the selection of banned $n$-grams is to filter the resulting multiset from the previous steps. We removed all non-frequent or short $n$-grams and made a set: 
\begin{equation}
T = \{\forall g \in \prescript{1}{}{ \mathfrak{T}}: \mu_{\mathfrak{T}}(g) > k_{min} \vee |g| > l_{min} \}    
\end{equation}
where $\mu $ is a multiplicity function, $k_{min} = 5, l_{min} = 4$. Then we used a train collection: each element of the set $T$ was checked to see if it was responsible for the result when the algorithm false positively triggered on at least one element of the collection. If so, the element was removed from the set. The train collection contains 20000 messages and 100\% of them are negatively labeled. 

\subsection{Specifics of moderating in real chat room}

After deploying the first version of our solution in a production run, we found that despite the low false positive rates the actual number of reported chat sessions with false positive errors was about 1.8 times higher than FPR. When working with real chat rooms, it should be kept in mind that counting moderation quality metrics on messages does not reflect the real user experience. The user does not count metrics on messages, but evaluates the whole interaction session with the LLM. Even one false positive evaluation of a message spoils the interaction experience for the whole session. In order to estimate the probability of unsuccessful session for a user, we used Bernoulli's formula:
\begin{equation}
P_{t} = 1 - (1 - FPR)^t    
\end{equation}
$FPR$ is a false positive rate of the moderation by messages, $P$ is a probability of at least one false positive moderator activation during a session of $t$ messages length. It is not hard to estimate that for a chat of length 5 messages, with our initial $FPR$ of 1\%, we get a 4.9\% probability of false moderator activation. If, however, we check both the user message and the model response, rather than just the model response, the probability of an undesirable outcome increases to 9.5\%. In practice, five and ten times the number of false positive errors is not achieved because inference sampling is heavily biased towards safe use of the dialogue system. However, this means that in real chat rooms, one has to be very careful with moderation; otherwise, the number of law-abiding users affected by overly active moderation may exceed acceptable limits \cite{Schwemer2024}. 

\subsection{Implementation details}

The computational complexity of the characteristic function of two sets depends linearly on the length of the shortest set. The set of forbidden $n$-grams contains about $10^5$ or more elements, while the number of words in the processed sequence is in a range of $10-1000$ words. The computational complexity of FLAME on inference depends linearly on the length of the model output. In production it takes 2 to 5 ms to check 1 message (4.3 ms on average at the real chat room) using only 0.1 CPU core and 100 Mb RAM.

\section{Results} 

Our test collection contains 9178 messages. The collection is balanced in terms of classes: 54\% of samples have a positive label. Metrics on the test collection are shown in Table \ref{tab:single}. 

\begin{table}[htbp]
    \centering
    \caption{\label{tab:single}Metrics of moderation quality by individual message}
        \begin{tabular}{ccccc}
         \hline
            \textbf{Precision} & \textbf{Recall} & \textbf{$F_1$} &  \textbf{FPR}&\textbf{Support} \\ \hline
            98.7$\pm$0.02\%&  90.9$\pm$0.04\%& 94.7$\pm$0.02\%&  1.38$\pm$0.02\%&9178\\ \hline
    \end{tabular}
\end{table}

Standard error for all metrics calculated by bootstrapping. The metrics shown in Table \ref{tab:single} are good enough to deploy the method in production. However, in order to verify the quality of the proposed solution, we conducted a series of experiments to compare the effectiveness in repelling the latest SOTA jailbreak on popular LLMs -- best-of-n (BoN) jailbreak \cite{hughes2024bestofnjailbreaking}. Enough time has passed since the original article with the BoN was released that the LLM bot holders have had time to issue some sort of response to it. Therefore, we also provide data on the current state of the moderation quality of APIs of various LLM chat bots. We used the methodology described in \cite{hughes2024bestofnjailbreaking} with one exception. We translated the dataset presented there into another language, which our solution was originally trained on. \\

The results of the experiments are presented in Figures \ref{fig:asr_final} and \ref{fig:fig1}. Table \ref{tab:final} presents the maximum achieved ASR values for each LLM with and without FLAME. FLAME worked relatively well for DeepSeek and ChatGPT, showing 9 times more effective resistance to attacks than the moderation system built into the their own API. In the case of DeepSeek, the BoN jailbreaking method achieved 100\% success, and quite quickly. Indeed, a check of the model output showed that their “constitution” of the moderation system differs significantly from the other models. DeepSeek comfortably chats about sensitive political topics as long as they do not involve recent Chinese history. The worst performance was shown for Claude -- jailbreak achieved its goal even with the presence of FLAME in almost half the cases which is more than for any other LLM. Taking into account their recent article \cite{sharma2025constitutionalclassifiers}, Anthropic appear to be in the process of redesigning their moderation system. \\

The best absolute performance was shown for GigaChat. We also compared the quality of the built-in GigaChat moderation system with the pure FLAME. Indeed, combining moderation systems gives a slightly higher result in resistance to attacks. However, in absolute terms it is insignificant (see Figure \ref{fig:fig1}). The effect of the accumulation of the probability of false positive errors during the chat session described in the section above makes this idea very risky in terms of the quality of the user experience.\\

\begin{table}[htbp]
    \centering
    \caption{\label{tab:final} Comparison of the BoN attack success rate (ASR) on moderation systems of different LLMs with that of FLAME}
        \begin{tabular}{cccc}
         \hline
            \textbf{Model} & \textbf{Base ASR}& \textbf{FLAME ASR} &  \textbf{B/F ASR ratio} \\ \hline
            GigaChat Max &  27.7$\pm$2.2\%& 3.8$\pm$0.8\% &  7.3 \\ \hline
            DeepSeek v3  &  100.0$\pm$3.0\%& 11.3$\pm$1.3\% &  8.9 \\ \hline
            Claude 3.5 Haiku &  93.1$\pm$3.3\%& 47.2$\pm$2.9\% &  2.0 \\ \hline
            Gemini Flash 1.5 &  44.7$\pm$3.1\% & 11.3$\pm$1.6\% &  4.0 \\ \hline
            GPT-4o-mini 8B &  77.4$\pm$3.6\% & 8.2$\pm$1.0\% &  9.4 \\ \hline
            Llama 3 8B &  97.5$\pm$4.4\% & 29.6$\pm$2.3\% & 3.3  \\ \hline
        \end{tabular}
\end{table}

\section{Discussion} 

Our experimental results reveal several critical insights about the current state and future directions of LLM content moderation. First, the varying effectiveness of FLAME across different models illuminates important patterns in moderation system design. The superior performance with GigaChat demonstrates the value of model-specific training, while the challenges faced with Claude (47.2\text{\%} attack success rate) highlight how architectural differences in LLMs can impact moderation effectiveness. \\

The difference between the result shown by FLAME for GigaChat and for the other models (see Figure \ref{fig:asr_final}) underscores the importance of knowing which model the engine is trained on and which one it infers. It is not overly surprising that FLAME showed the best quality on the same model, whose answers were used to train it. \\
The real-world deployment of FLAME has provided valuable insights into practical implementation challenges. One observation is that the false positive rates in production environments can be 1.8 times higher than in isolated testing, emphasizing the vital role of considering complete user sessions rather than individual interactions. This finding fundamentally changes how we should approach evaluation of moderation systems and their optimization and should be a subject of a benchmarking effort in future work.\\

Analysis of combined moderation approaches yielded unexpected insights. While integrating FLAME with existing systems showed marginal improvements in attack resistance, the multiplicative effect on false positives suggests that simpler, focused approaches may be more efficient in practice. This challenges the common assumption that layering multiple security measures necessarily improves overall system safety.\\
The production deployment also revealed interesting patterns in user interaction and system performance under real-world conditions. The relationship between chat session length and cumulative false positive rates provides a recipe on how moderation systems should be calibrated for different use cases. These insights extend beyond FLAME's specific implementation and are positioned to influence broader design principles of moderation systems.

\subsection{Limitations} 

FLAME is inexpensive to train and infer, showing acceptable quality on the test sample, and is highly resistant to the SOTA attack method. However, it also has limitations. Firstly, its performance results strongly depend on the difference between the model used during training and the one that will be used in inference. One requires the engine to be trained separately on a model on which it will be used. Secondly, its training requires access to an unmoderated version of the model, which are not always available.

\section{Conclusion} 
FLAME represents a significant advancement in LLM content moderation, demonstrating that efficient protection against modern jailbreaking techniques can be achieved through lightweight but powerful approaches. Our comprehensive evaluation across multiple leading LLM platforms shows that FLAME consistently reduces attack success rates by a factor of 2--9 compared to existing solutions, while maintaining minimal computational requirements of just 0.1 CPU core and 100 MB RAM per instance.
The system's success in real-world deployment validates our approach of shifting focus from input filtering to output moderation. This paradigm shift, combined with our rule-based architecture enhanced by LLM-generated training data, challenges the prevailing trend toward increasingly complex and resource-intensive censorships. The results demonstrate that successful moderation can be achieved without extensive model fine-tuning or complex neural architectures.
Our work establishes a new direction for developing practical, scalable content moderation systems, protecting against adversarial attacks and providing computational efficiency and a true flexibility in deployment. As LLMs continue to evolve and integrate into all sorts of applications, approaches like FLAME will be crucial in ensuring safe and appropriate human-AI interactions.

\bibliographystyle{ACM-Reference-Format}
\bibliography{literature}



\end{document}